\documentstyle[12pt,aaspp4]{article}

\begin{document}

\title{Selection Effects, Biases, and Constraints in the Cal\'{a}n/Tololo Supernova Survey}

\author{Mario Hamuy and Philip A. Pinto\altaffilmark{1}}
\affil{University of Arizona, Steward Observatory, Tucson, Arizona,
  85721; email: mhamuy@as.arizona.edu, ppinto@as.arizona.edu}
\altaffiltext{1}{Lawrence Livermore National 
Laboratory, Livermore CA 94550 USA}

\begin{abstract}

We use Monte Carlo simulations of the Cal\'{a}n/Tololo photographic
supernova survey to show that a simple model of the survey's selection
effects accounts for the observed distributions of recession velocity,
apparent magnitude, angular offset, and projected radial distance between
the supernova and the host galaxy nucleus for this sample of Type Ia
supernov\ae (SNe~Ia). The model includes biases due to the flux-limited
nature of the survey, the different light curve morphologies displayed by
different SNe~Ia, and the difficulty of finding events projected near the
central regions of the host galaxies.  From these simulations we estimate
the bias in the zero-point and slope of the absolute magnitude-decline rate
relation used in SNe~Ia distance measurements. For an assumed intrinsic
scatter of 0.15 mag about this relation, these selection effects decrease
the zero-point by 0.04 mag. The slope of the relation is not
significantly biased. We conclude that despite selection effects in the
survey, the shape and zero-point of the relation determined from the
Cal\'{a}n/Tololo sample are quite reliable. We estimate the degree of
incompleteness of the survey as a function of decline rate and
estimate a corrected luminosity function for SNe~Ia in which the
frequency of SNe appears to increase with decline rate (the fainter SNe are
more common). Finally, we compute the integrated detection efficiency of
the survey in order to infer the rate of SNe~Ia from the 31 events
found. For a value of H$_{0}$=65 km sec$^{-1}$ Mpc$^{-1}$ we obtain a SN~Ia
rate of 0.21$^{\rm +0.30}_{-0.13}$ SNu. This is in good agreement with the
value 0.16$\pm$0.05 SNu recently determined by \cite{capellaro97}.

\end{abstract}

\keywords{statistics --- supernovae: general}

\section{Introduction}

The Cal\'{a}n/Tololo (CT, hereafter) Survey was a classical photographic
search for supernovae initiated in June 1990 with the principal goal of
examining the Hubble diagram for Type Ia supernovae (SNe~Ia) out to
redshifts of $\sim$0.1 (\cite{hamuy93}, hereafter referred to as Paper I).
Upon its completion in November 1993, the survey had discovered 49 SNe, 31
of which were spectroscopically classified as SNe Ia (\cite{hamuy96c},
hereafter referred to as Paper VII).  Because the survey was carried out in
a systematic and uniform way and complemented with followup photometric and
spectroscopic observations of unprecedented quality, it provides a unique
opportunity to study the properties of the Type Ia family.

In any survey, however, selection effects present a major problem in
drawing firm conclusions from the observed data. Because the CT survey was
flux-limited, we expect that Malmquist bias will have favored the discovery
of the intrinsically-brightest objects of the supernova luminosity
distribution and suppressed the detection of the intrinsically-faintest
events over most of the volume surveyed.

Another bias is expected which arises from the different rates of evolution
exhibited by SNe~Ia.  It is by now well established that the peak
magnitudes of SNe~Ia are correlated with the duration of the peak in their
lightcurves (\cite{phillips93}, \cite{hamuy96a}, hereafter referred to as
Paper V). Because the survey was limited to relatively infrequent
observations when compared with the length of time the SNe are at their
brightest, a selection effect similar to Malmquist bias will occur in which
the bright supernov\ae\ are more likely to be detected not only because
they are brighter but also because they last longer.

A possible consequence of these biases is a shift of the zero-point of the
Hubble diagram, and hence the Hubble constant, to larger values.  Because
SNe~Ia are seen to exhibit a large scatter ($\sim$2 mag in B) in their
maximum-light luminosities (\cite{phillips93}, Paper V), this shift could
be substantial. Using the peak luminosity-decline rate relation,
initially found by \cite{phillips93} from nearby SNe~Ia, the observed peak
magnitudes can be ``corrected'' to produce a Hubble diagram with
significantly lower scatter $\sim$0.10-0.15 mag (\cite{hamuy96b}, hereafter
referred to as Paper VI; see also \cite{riess96} for a different approach
to applying the same correction). Although we can expect the bias in the
zero-point of this corrected Hubble diagram to be significantly reduced,
concern has been expressed regarding the actual shape of the peak
luminosity-decline rate relation (hereafter referred to as
M/$\Delta$m$_{15}$ relation) which determines the amount of correction to
be applied to the observed magnitudes.  In fact, the relation
determined from the CT sample appears to be shallower than that implied by
the nearby SNe (Paper V), a possible consequence of selection effects in
the photographic survey.

Both the zero-point and slope of the M/$\Delta$m$_{15}$ relation are
relevant to the determination of the Hubble constant from
intermediate-distance SN samples and of cosmological parameters (such as
$\Omega_{M}$ and $\Omega_{\Lambda}$) from high-redshift SNe~Ia
(\cite{perlmutter95}, \cite{schmidt98}).  The very existence of such a
correlation, and its detailed form, is an important clue to understanding
the nature of the explosions themselves.

The main goal of this work is to derive a simple model for the selection
effects of the CT survey and for the parent population of SNe~Ia that
suitably reproduces the observational properties of the $\sim$30 discovered
SNe~Ia.  This model may then be used to estimate the biases in the
observational parameters derived from the CT sample and to infer intrinsic
properties of the Ia family like the M/$\Delta$m$_{15}$ relation and the
luminosity function.

After summarizing the observational data (Section 2), in Section 3 we use
simulations to test the selection effects proposed and derive parameters
which best fit a model for the parent population based on the {\it
observed} M/$\Delta$m$_{15}$ relation and a {\it flat} luminosity
function. After exploring its parameter space in some detail, in Section 4
we use this model to estimate the biases in the observed
M/$\Delta$m$_{15}$, luminosity function, and galaxy type-redshift relation.
With {\it corrected} versions of these functions, in Section 5 we iterate
our simulations in order to test our revised assumptions about the
intrinsic properties of SNe~Ia.  Armed with this improved model, in Section
6 we compute the detection efficiency of the survey and estimate the
actual rate of SNe Ia.

\section{Observations}

The details of the CT Supernova Survey were extensively described in Paper
I. In this section we summarize only those features of the survey necessary
to modeling selection effects.  The search phase of the survey was
begun in June 1990 using the Cerro Tololo Interamerican Observatory (CTIO)
Curtis Schmidt Camera which has a 19.05 cm x 19.05 cm field and a scale of
96.6 arcsec mm$^{-1}$, providing a useful sky coverage of 26.13 square
degrees when used with photographic plates. The search was performed with
hypersensitized (unfiltered) IIa-O plates which yielded a limiting
magnitude for isolated sources of m$_{pg}$ $\sim19$ with an exposure of 20
min on moonless nights. The plate scanning was performed by experienced
assistants at the Department of Astronomy of the University of Chile.  A
detailed comparison of each plate with a first-epoch exposure was carried
out visually with a Zeiss-Jena blink comparator.

The survey began with observations of 60 fields located at high galactic
latitudes covering from 0 to 24 h in right ascension. Given the large area
covered by the photographic plates, the resulting sample of galaxies
surveyed in the program was not seriously biased to any particular Hubble
type.  Table 1 of Paper I lists the equatorial coordinates of the selected
fields.  During 1990 and 1991 each field was observed at $\sim$30 day
intervals. As of March 1992 the frequency of observations was increased to
twice per month in order to discover SNe at earlier stages in their
evolution. This required the elimination of 15 fields from the initial
list; the remaining 45 fields were observed until the end of the search
phase on November 1993. A total of 1,019 plates were obtained in the course
of the program. Figure \ref{f_obstime} shows the complete time sampling of
the 60 fields.

The survey discovered a total of 49 SNe, of which 31 were spectroscopically
confirmed as SNe~Ia \footnote[1]{Note that in previous papers of this
series an additional SN~Ia (1992au) was included in the list of
discoveries. This will not be considered in the statistical analysis of
this paper since this object was found as part of a QSO search on an
objective prism IIIaJ plate.}.  Table 1 gives a complete listing of the
observed properties of these SNe~Ia, of which useful follow-up CCD
photometry was obtained for 26 events. These constitute our sample of
``best-observed'' SNe. 60\% of these occurred in spiral galaxies.

Some of the characteristics of the CT sample can be appreciated from the
histograms shown in Figure \ref{f_ctdata}.  The recession velocity
distribution shows a systematic increase in the number of supernov\ae\
detected in redshift bins with velocities from 0 to $\sim$15,000 km
sec$^{-1}$. This is expected from a spatially uniform distribution of SNe
as the volume of each velocity (distance) bin ($4\pi r^{2}{\it d}r$)
increases quadratically with distance. At larger distances the SNe become
significantly harder to discover because they appear fainter to the
observer, the objects remain above the detection threshold for a shorter
time, and a larger fraction of the SNe appear projected against the
over-exposed central parts of the host galaxies (the ``Shaw'' effect
described below).  The cutoff at $\sim$30,000 km sec$^{-1}$ is the
signature of a flux-limited survey; no supernovae are bright enough to be
detected at greater distances with the survey's nominal limiting magnitude.
Given the linear relation of the Hubble law between apparent magnitude and
the logarithm of the redshift, the peak apparent magnitude distribution is
similar to the velocity histogram, reaching a maximum at B$\sim$17.5.

The absolute magnitude distribution displays a sharp peak around
$M^{B}_{MAX}=-19.0+5log(H_{0}/65)$, with a dramatic decline toward less
luminous events.  It is hard to ascertain at this stage whether the cutoff
is real or due to statistical fluctuations arising from the small sample
size.  Clearly, the overall shape of this distribution does not correspond
to the actual luminosity function of SNe~Ia as intrinsically fainter events
are less likely to be discovered by the search.  The decline rate
distribution displays a similar behavior to that of absolute magnitude, in
accord with the linear relation between peak luminosity and decline
rate displayed by the CT SNe (Paper V). A number of differences are
evident, however.  The decline rate distribution is somewhat flatter, with
a shallow maximum around $\Delta$m$_{15}$(B)=1.2 with no obvious cutoff
toward fast-declining (faint) SNe.

The angular separation histogram has a sharp maximum at $\sim$7 arcsec and
trails off toward larger offsets.  The remarkable lack of objects discovered
at projected angular separations less than 4 arcsec is a clear indication
that the search fails to discover SNe at or near the center of the surveyed
galaxies -- it is difficult to imagine a physical effect which would lead
to the absence of supernov\ae\ where the stellar density is greatest.  This
effect was noted originally by \cite{shaw79}, who interpreted it as an
observational bias inherent to all photographic surveys; the central parts
of the parent galaxies are over-exposed, making it impossible to detect the
additional brightening from a supernova.  The projected radial distance
distribution is similar to the angular separation histogram. The
distribution has a peak at $\sim$6 kpc followed by a tail suggestive of
the distribution of light from an exponential disk.

The ideal tool for studying the spatial distribution of the discovered
events is the V/V$_{MAX}$ test. Unfortunately, this requires an estimate of
the limiting magnitude at which supernov\ae\ could be discovered for each
of the discovery plates.  While it is in principle possible to measure the
limiting magnitude for isolated sources for each observation, the threshold
for SN detection is much harder to estimate from photographic plates since
the SN are generally superimposed on the bright background of the host
galaxies and a plethora of effects such as confusion and image saturation
make the SN detection threshold a function of position and host galaxy
morphology as well as the overall plate limit.  We have therefore been
unable to calculate V/V$_{MAX}$ values for the CT SNe.

\section{Simulating the Survey}

Our goal is to propose a simple model for the selection effects of the CT
photographic survey which employs the fewest assumptions and parameters yet
accounts as closely as possible for the observed distributions in Figure
\ref{f_ctdata}. To test the model, we have performed Monte Carlo
simulations of a parent population of SNe with the following
characteristics.

\subsection{Sampling from the parent population}

We assume that the SNe occur in galaxies. Although the galaxy
distribution is clumpy, the CT survey covers a sufficient volume of space
that ``clumpiness'' from large scale structure should not have a
significant effect. Each photographic plate covers $\sim$ 230
(H${_0}$/65)$^{-1}$ Mpc in depth and $\sim$ 20 (H${_0}$/65)$^{-1}$ Mpc on
the plane of the sky (for an average redshift of 15,000 km
sec$^{-1}$). This is several times larger than the galaxy clustering length
of 8.3 (H${_0}$/65)$^{-1}$ Mpc obtained from analysis of the angular
two-point correlation function (\cite{peebles93}).  We therefore assume for
the purpose of this investigation that the parent population of SNe is
uniformly distributed in space, and to each generated SN we randomly assign
a distance within a spherical volume. The radius of this sphere is chosen
to be just larger than the maximum distance to which the most luminous
event can be observed (given the limiting magnitude of the simulation).  We
parametrize the time of the SN explosion by the time of maximum light which
we randomly sample within a period of 1,460 days, the 4 years during
which observations were obtained. Because it is potentially possible to
discover supernovae at times much later than maximum light, this time
window begins on JD 2447866.5 which corresponds to 200 days before the
first observation epoch of the CT survey.  Because the risetime of the B
lightcurve is $\lesssim 18$days, the window ends 19 days after the last
observing run.

To each generated SN we assign an absolute magnitude from the
M/$\Delta$m$_{15}$ relation found in Paper V, namely,
\begin{equation}
M^{\rm B}_{MAX} - 5 log(H_0/65) = a_0 + b_0 [\Delta m_{15}({\rm B}) - 1.1],
\label{eqn1}
\end{equation}
where a$_{0}$=-19.258 and b$_{0}$=0.784 \footnote[2] {Note that in Paper V
(and the other papers of the series) the labels of the figures have the
wrong sign for the logarithm. This is a purely cosmetic error which affects
neither the content nor the conclusions of the papers.}.  Note that both of
these coefficients are entirely determined from the observed sample of
SNe~Ia and do not depend on the choice of the Hubble constant
(H$_{0}$). The latter can be obtained from a$_{0}$, b$_{0}$, and the
absolute magnitude of a SN~Ia with known decline rate.  If M$^{\rm
B}_{1.1}$ denotes the absolute magnitude of a ``standard'' SN~Ia, one
corrected to a decline rate of $\Delta m_{15}(\rm{B})=1.1$, equation
(\ref{eqn1}) becomes,
\begin{equation}
M^{\rm B}_{1.1} - 5 log(H_0/65) = a_0,
\label{eqn2}
\end{equation}
from which the value of H$_0$ can be obtained directly. In fact, this
method was employed in Paper VI using the absolute magnitude of nearby
SNe~Ia recently determined by Sandage and collaborators (\cite{sandage96},
and references therein) to infer a value of H$_0$(B)$\sim$64. For the sake
of generality, we have chosen in this paper to leave the Hubble constant as
a free parameter in the simulations, parameterizing it with M$^{\rm
B}_{1.1}$ according to equation (\ref{eqn2}).

There are several potential problems in adopting the observed
M/$\Delta$m$_{15}$ relation.  First, this equation was obtained from a
sub-sample of CT SNe with decline rates ranging between
0.87$\leq\Delta$m$_{15}$(B)$\leq$1.7.  At decline rates larger than 1.7,
the CT sample contains only one object, SN 1992K, with
$\Delta$m$_{15}$(B)=1.93, which appears to be 2$\sigma$ fainter than the
value given by equation (\ref{eqn1}).  Second, it may well be that the
actual shape of this function departs from the assumed linear form, even on
the range 0.87$\leq\Delta$m$_{15}$(B)$\leq$1.7. A recent re-examination of
the M/$\Delta$m$_{15}$ relation after applying corrections for host galaxy
extinction suggests that the data support fitting this relation by a
quadratic function (\cite{phillips98}).  In any case, as the underlying
cause of the relation is largely unexplained, there is no physical reason
to expect {\em any} particular functional form. A linear fit is an adequate
approximation for the majority of objects observed and a convenient choice
for the purpose of illustrating selection effects.  In section 4.1 we will
consider simulations with a quadratic function. The third possible problem
with equation (\ref{eqn1}) is that the coefficients a$_0$=-19.258 and
b$_0$=0.784 could be substantially biased by the selection effects of the
search. The assumption of a M/$\Delta$m$_{15}$ relation of the form of
equation (\ref{eqn1}) for the parent population will be reviewed in detail
in Section 4.1, after evaluating possible observational biases in the
search.

The process of assigning an absolute magnitude to each generated SN starts
by generating a uniform random value for the decline rate in the interval
$\Delta$m$_{15}$(B)=0.87-1.93 which spans the entire range of decline rates
of the best-observed SNe~Ia. By construction, the decline rates of the CT
SNe were also limited to this interval from comparison of the light curves
of these objects to that of a family of six template lightcurves where SNe
1992bc (with $\Delta$m$_{15}$(B)=0.87) and 1991bg
($\Delta$m$_{15}$(B)=1.93) are at the extremes of the distribution
(\cite{hamuy96d}, hereafter Paper VIII).  The assumption of a flat decline
rate distribution in the simulation (which, since the relation between
absolute magnitude and decline rate is assumed to be linear, corresponds to
a flat luminosity function for SNe~Ia) is just a hypothesis at this stage,
one which is further examined in Section 4.2. Given the decline rate, we
use equation (\ref{eqn1}) to calculate the SN absolute magnitude.  To this
we add a Gaussian scatter with $\sigma$$_{M}$=0.15 mag (consistent with the
observational scatter of the corrected Hubble diagram).  The distance and
absolute magnitude are then used to calculate the peak apparent
magnitude. The recession velocity is determined from Hubble's law, the
chosen value of the Hubble constant (computed from the absolute magnitude
M$^{\rm B}_{1.1}$), and an additional Gaussian scatter of $\sigma_{V}$=600
km sec$^{-1}$ to allow for the peculiar velocity of the host galaxy.

\subsection{Adding the selection effects}

We define a successful SN discovery as an event which satisfies two
criteria. The first is that at the time of observation the event has an
apparent magnitude brighter than the limiting magnitude of the photographic
plate. The second is that it has a minimum projected separation from the
nucleus of the parent galaxy.

Because a SN is a transient event we not only need to know its peak
apparent magnitude but also its brightness at the specific time of the
observation. This requires us to model the luminosity evolution of events
which have been found to exhibit a wide range of light curve
morphologies. In Paper VIII we presented a family of six template light
curves for SNe~Ia for days (relative to B maximum) -5 through +80 and
spanning the entire range of observed decline rates of the best-observed
SNe~Ia ($\Delta$m$_{15}$(B)=0.87-1.93).  In this paper we adopt an expanded
version of these templates which covers a wider range in age (from day -14
to +600).  The new values, given in Table 2, were obtained from photometry
of these same six SNe. In those cases where the photometric data were
insufficient to span the desired age range, it was necessary to extrapolate
the new values.  The extrapolations to late epochs are fairly certain since
SNe display a linear decline in magnitude after day $\sim$30;
as shown in Section 5 no supernovae were ``discovered'' in the simulations
after day 255.  Before maximum light, on the other hand, the templates are
much steeper and the extrapolation suffers an additional uncertainty.

Armed with this family of six extended B templates, for each simulated SN
we perform an interpolation in decline rate in order to produce a synthetic
B light curve. Then we randomly select one of the CT field numbers in order
to choose a time sampling (this does not constrain the SN
candidate by its location in the sky).  For the first observation epoch of
the selected field (which is taken from the observing records of the
survey) we calculate the apparent magnitude of the generated SN. Since the
age of the SN at the specific observation epoch is known (from the randomly
generated epoch of maximum and the time of observation), we interpolate in
time within the synthetic light curve in order to calculate the magnitude
difference of the SN between the observation epoch and peak. Because we
know the peak apparent magnitude, we may finally calculate the
present-epoch apparent magnitude and ask whether the SN is brighter than
the limiting magnitude.  If the SN is detected on the first plate,
we add this object to the list of positive detections and move on to the
next SN candidate.  If the SN is not detected in the first epoch we
continue with the subsequent epochs of observation for the selected field
until the SN is either detected or the survey of that field is terminated.

Although we know the nominal limiting magnitude of the survey
(m$_{pg}\sim$19), the individual values for the complete set of plates are
not available nor do we know the limiting magnitude for SN discovery. Thus,
in the simulations the limiting magnitude is a free parameter (B$_{lim}$)
with a characteristic fluctuation of $\sigma_{Blim}$=0.3 mag. This
fluctuation allows for variations of the detection threshold between plates
due to seeing, clouds, the phase of the Moon, plate quality, the difficulty
of finding SNe projected on galaxies with different surface brightnesses,
etc.

As we do not have light curve information before day -14, we assume that
SNe younger than this age are not detected, and to increase the efficiency
of the simulation we assume that SNe older than 600 days are too faint ever
to be seen.  These assumptions have very little effect on our results since
only a very small fraction of the generated SNe are located close enough to
the observer to be brighter than the detection threshold for ages outside
these limits.

As seen in the angular separation histogram of Figure \ref{f_ctdata}, the
survey is clearly biased against the detection of SNe near the center of
the host galaxy. This occurs because such SNe have low contrast with
respect to the bright background (frequently overexposed) on which they
lie, increasing the difficulty of discovery in a visual scan of the
plate. For our simple model we assume that the SNe occur in idealized host
galaxies consisting of smooth exponential disks characterized by a
luminosity scale length (a free parameter in the simulations which we
denote r$_0$=1/$\alpha$), with the radial distribution of SNe following
that of the host galaxy light.  The radial distribution is assumed to be
the same for all SN~Ia lightcurve morphologies.  We assign random
orientations to the disk galaxies with respect to the observer in order to
calculate the projected angular offset. To account for the observed bias, a
successful detection requires not only an apparent magnitude brighter than
the limiting magnitude but also a minimum projected separation between the
SN and the parent galaxy center of 4 arcsec (in agreement with the observed
limit from Figure \ref{f_ctdata}).

We have made the computationally expedient choice to represent all host
galaxies as exponential disks. Although a significant number of events
(40\% of the total) occured in ellipticals (see Table 1), this choice
affects only the estimate of the SN rate (section 6). Given the other,
larger uncertainties involved in this calculation such a simplification has
only a marginal effect.  When appropriate, in the following sections we
will include in our discussion radial distributions following a de
Vaucouleurs r$^{1/4}$ law characterized by an effective radius r$_{eff}$.
As will be seen, the effect of changing the form of the radial distribution
we adopt is small.

Of course, the real detection probability depends on many other factors
such as the proximity to spiral arms or H$_{II}$ regions (confusion), or
the presence of possibly strong and patchy obscuration by dust. We do not
consider any such further effects as they are difficult to model without
introducing numerous ill-defined free parameters. We shall find below that
our simple model, without such embellishments, is quite sufficient to
reproduce the observations.

\subsection{Exploring the parameter space}

The results of the simulations depend upon a number of parameters which are
listed in Table 3. While the determination of all of these may, in
principle, be biased by the manner in which the survey was conducted, most
either do not affect the results to any significant extent or are fixed to
reasonable accuracy by observation. a$_0$, b$_0$, and $\sigma_M$ have been
determined from the survey in Paper V. The choice of peculiar velocity
dispersion ($\sigma$$_{V}$) is well determined by observations
(\cite{marzke95}) and proves to be a small fraction of the recession
velocity for most of the discovered SNe. The value of the scatter in the
limiting magnitude ($\sigma_{Blim}$) is just a guess at this stage. The
value of the angular separation cutoff seems well determined by the
histogram in Figure \ref{f_ctdata}.  We will adopt these values (denoted
``fixed'' in Table 3) for present purposes and their sensitivity to
selection effects will be explored in Section 4 (a$_0$, b$_0$, and
$\sigma_M$) and Section 5 ($\sigma_{Blim}$).

To begin our exploration of this space of parameters, we choose to vary the
three most important as determined by numerical experiment: M$^{\rm
B}_{1.1}$, B$_{lim}$, and $\alpha$.  We divide the range of plausible
values for these parameters, -19.758$\leq$M$^{\rm B}_{1.1}\leq$-17.958,
17.8$\leq$B$_{lim}\leq$18.8, and 0.05$\leq\alpha\leq$0.50 kpc$^{-1}$, into
ten intervals each and perform simulations for each of the 1000
combinations, stopping the simulation after 9,000 ``supernov\ae'' are
detected.

We measure the relative merit of a given simulation by comparing the
generated histograms in recession velocity and angular separation to those
of the survey given in Figure \ref{f_ctdata}. The probability that a given
observed histogram is drawn from the same distribution as the synthetic
data is calculated from a Kolmogorov-Smirnov (KS) statistic, and the
overall success is measured by the joint probability that both
distributions match.  A first check on the model is that a satisfactory
solution exists for some combination of these parameters.

Figure \ref{f_simpcont} shows the resulting contour levels of the KS
probability for the angular separation distribution (thin lines) in the
B$_{lim}$-$\alpha$ plane, for four different values of M$^{\rm B}_{1.1}$. It is
remarkable that these surfaces are well organized and exhibit
simply-connected maxima within the adopted parameter range, and further
remarkable that the maxima for the different probabilities overlap.

For a fixed value of M$^{\rm B}_{1.1}$, the KS probabilities for the
angular separations (thin lines) are most strongly affected by $\alpha$
which measures the scale length of the exponential disks. The probabilities
prove quite insensitive to the limiting magnitude of the simulation. It is
interesting to note that the value of $\alpha$ that yields the highest KS
probability for the angular separation distribution increases for fainter
SN absolute magnitudes (i.e the scale length decreases for fainter absolute
magnitudes).  This comes about because fainter absolute magnitudes imply
that successful detections will occur in galaxies located at smaller
distances. This is shown in the left panels of Figure \ref{f_fourm} for two
widely different values of M$^{\rm B}_{1.1}$.  Smaller values of M$^{\rm
B}_{1.1}$ make the characteristic angular size of the SN host galaxies
increase, broadening the distribution of projected angular separations as
can be seen in Figure \ref{f_fourm} (right panels).  The only way to
compensate for this broadening and simultaneously match the observed
distribution of angular separations (Figure \ref{f_ctdata}) is to decrease
the scale length of the disks (by increasing $\alpha$).  What we learn from
this test is that different combinations of $\alpha$ and M$^{\rm B}_{1.1}$
values can give equally good KS probabilities for the angular separation;
it is not possible to break this degeneracy with this test alone.

Returning to Figure \ref{f_simpcont}, the thick lines show the contour
levels of the KS probability for the recession velocity distributions.  For
a fixed value of M$^{\rm B}_{1.1}$, these probabilities are quite sensitive
to B$_{lim}$. This is to be expected since the upper cutoff of the
recession velocity distribution of SNe is determined by the limiting
magnitude for SN detection. These probabilities depend somewhat on $\alpha$
but very little on M$^{\rm B}_{1.1}$. For a given value of M$^{\rm
B}_{1.1}$ it is possible to identify in this diagram the values of the
$\alpha$ and B$_{lim}$ parameters which {\it simultaneously} yield the
highest KS probabilities for the angular separation and recession velocity
distributions.  These ``best-values'' are indicated with the $\oplus$
symbol in Figure \ref{f_simpcont}. For B$_{lim}$, we find a ``best-value''
of 18.3 for all values of M$^{\rm B}_{1.1}$. For $\alpha$ we find a
``best-value'' between 0.15 and 0.3 kpc$^{-1}$, depending on M$^{\rm
B}_{1.1}$.

In order to examine the possibility of a bias in the determination of
$\alpha$ from the inclusion of SNe in elliptical galaxies, we have
explored the same parameter space with elliptical hosts removed from the CT
sample (13 events).  Although elliptical hosts dominate the events at large
radial separations, the distributions of angular separation for ellipticals
and spirals are very similar (see Figure \ref{f_ctdata}). We thus do not
expect the value of $\alpha$ we obtain to be sensitive to the inclusion of
SNe with elliptical hosts and, indeed, the simulations confirm this.
Another interesting test is to determine the effective radius for
elliptical hosts employing a SN radial distribution following an r$^{1/4}$
law. An examination of the parameter space determined by the distribution
of angular separations for the 9 CT events in early type galaxies yields a
``best-value'' for r$_{eff}$ between 3 and 7 kpc, depending on M$^{\rm
B}_{1.1}$ (r$_{eff}$ decreases for fainter absolute magnitudes).  For
B$_{lim}$ we find again a ``best-value'' of 18.3.

The ``best-value'' of 18.3 for B$_{lim}$ is reasonable since the nominal
limiting magnitude of the plates for isolated sources is $\sim$19 and the
limiting magnitude for SN discovery must be brighter than this given the
additional difficulties in finding SNe described above.  If we
pick an absolute magnitude consistent with that of Paper VI (M$^{\rm
B}_{1.1}$=-19.258; H$_{0}$=65), the ``best value'' for $\alpha$ is 0.15
kpc$^{-1}$, which corresponds to a scale length of r$_0$=6.7 kpc for the
disks of the host galaxies. This value lies within the range (0.6-8.0 kpc)
of observed scale lengths for spiral galaxies, properly scaled to
H$_{0}$=65 (\cite{kruit88}).  Likewise, the ``best value'' for r$_{eff}$ of
6 kpc (H$_{0}$=65) proves to be in good agreement with those typically
measured in ellipticals (\cite{kormendy77}).

Figure \ref{f_simpcont} shows that the KS tests provide a means of
assessing the goodness of fit for each simulation and determining the best
values of $\alpha$ and B$_{lim}$.  The value of M$^{\rm B}_{1.1}$ (and thus
H$_{0}$) remains undetermined from this test, however. To illustrate the
agreement of this simple model with observations, we have run a simulation
with an absolute magnitude consistent with that of Paper VI and the
corresponding ``best-values'' B$_{lim}$=18.3 and $\alpha$=0.15
kpc$^{-1}$. Figure \ref{f_simphist} shows the recession velocity, apparent
magnitude, angular separation, and radial distance distributions of 50,000
detected SNe, along with the observed histograms.  The match to the
observations is satisfactory with KS probabilities of 0.74 and 0.91 for the
recession velocity and angular separation distributions, respectively. The
KS probabilities for the apparent magnitude and radial distance
distributions (0.46 and 0.20), on the other hand, are significantly lower,
but this is not unexpected since the ``best values'' for the fitting
parameters correspond only to those that maximize the KS probabilities for
recession velocity and angular separation.  For the time being we consider
this to be a satisfactory solution that will allow us to estimate biases in
the CT survey in the following section.

\section{Evaluation of Biases in the Cal\'{a}n/Tololo Sample}

We have shown that our simple model for the selection effects can
satisfactorily reproduce the angular separation and recession velocity
distributions of the CT SN sample.  In this section we employ the
``best-values'' for the model parameters (M$^{\rm B}_{1.1}$=-19.258,
B$_{lim}$=18.3, $\alpha$=0.15 kpc$^{-1}$) in order to quantify biases in
the {\it observed} M/$\Delta$m$_{15}$ and luminosity functions, and to look
for bias in the galaxy type-redshift relation.

\subsection{The absolute magnitude-decline rate relation}

We have assumed thus far that the parent population of SNe obeys the
M/$\Delta$m$_{15}$ relation given by equation (\ref{eqn1}) and have
ignored possible biases in the observed zero-point and slope of this
relation due to selection effects. In order to examine the effect that
the selection criteria have on constraining the parent population of SNe,
we first analyze an idealized situation where we suppress the peculiar
velocities of the host galaxies ($\sigma_{V}$=0), the scatter in limiting
magnitude ($\sigma_{Blim}$=0), and any scatter about the M/$\Delta$m$_{15}$
relation ($\sigma_{M}$=0). With these (over)simplifications we consider the
following four cases:

\noindent $\bullet$ Case I (Malmquist bias): We assume that the generated
events have constant luminosities (independent of time) and we select them
solely on the basis that their apparent magnitudes are brighter than
B$_{lim}$.  The left panel of Figure \ref{f_vvmax} (top) shows the absolute
magnitudes of 1,000 detected SNe as a function of recession velocity.  This
is the classical Malmquist bias of a flux limited survey, in which the
faintest objects of the luminosity distribution are lost near the detection
limit.  The diagonal dashed line is the relation
M$^{B}_{MAX}$=B$_{lim}$-5log(v$_{MAX}$/H$_{0}$)-25, and is drawn to
indicate the maximum recession velocity to which each SN type can be
found. The horizontal solid line indicates the average absolute magnitude
of the parent population (M$^{B}_{MAX}$=-19.023) and the horizontal dashed
line corresponds to the average magnitude of the detected population
(M$^{B}_{MAX}$=-19.091); their difference is the bias introduced in the
zero-point of the M/$\Delta$m$_{15}$ relation.  In this case the bias
amounts to only 0.068 mag.  In general, its magnitude is an approximately
quadratic function of the spread of the absolute magnitudes of the parent
population.  The corresponding V/V$_{MAX}$ distribution for the detected
SNe is shown in the right panel of this figure (top). The histogram is
flat, with an average value of 0.5, the expected distribution of a
flux-limited sample drawn from a spatially uniform parent distribution.

\noindent $\bullet$ Case II (Malmquist bias and Shaw effect): Here, objects
are selected not only by demanding that their apparent magnitudes are
brighter than B$_{lim}$, but also that their angular separations from the
nucleus of the host galaxy are sufficiently great (the Shaw
effect). Although it proves difficult to see any additional effect
with respect to Case I in the left panel of Figure \ref{f_vvmax} (second
plot from top to bottom), it is evident that the V/V$_{MAX}$ histogram
(right panel) becomes somewhat skewed toward lower values, a clear
indication that more distant objects are more difficult to discover.  This
comes about because the minimum angular separation of 4 arcsec required for
detection maps into increasingly larger radii in the more distant
galaxies. Fewer of the SNe generated in distant galaxies are detected
because a larger fraction of the stellar mass lies within the ``exclusion''
region. The Shaw effect reduces the detection efficiency by
$\sim 24$\% with respect to Case I.

\noindent $\bullet$ Case III (Malmquist bias and ``light-curve'' bias): We
now include the fact that the luminosity of SNe~Ia evolves on a timescale
comparable to the interval between observations, but ignore the effect of
proximity to the host galaxy nucleus. We generate the lightcurves of the
supernov\ae\ from the templates as described above.  For successful
detection we demand that the apparent magnitude {\em at the time of
observation} is brighter than the detection threshold. In this case the
most distant objects prove more difficult to detect because they remain
above the detection threshold for a shorter interval in time.  The
V/V$_{MAX}$ distribution for the detected SNe is now heavily skewed toward
lower values (right panel; third plot of Figure \ref{f_vvmax}, from top to
bottom). This is also evident in the left panel where the concentration of
objects near the detection limit (diagonal line) is significantly lower
than in Case I.  Note also that this effect is much more severe for the
dimmer events; the more rapid evolution of the less luminous objects
decreases their probability of discovery.  The combined effect of these two
selection effects is to increase the average absolute magnitude of the
detected SNe by 0.094 mag with respect to the value of the parent
distribution.  The ``light-curve'' effect reduces the detection efficiency
by 85.5\% with respect to Case I for the particular time sampling of the CT
survey. For a survey which examined each field once each day, on the other
hand, this effect would be insignificant.

\noindent $\bullet$ Case IV (Malmquist bias, Shaw effect, and
``light-curve'' bias): Finally, we consider the combined effect of
proximity, correlation between luminosity and duration, and limiting
magnitude.  As expected, the V/V$_{MAX}$ histogram becomes even more skewed
toward lower values (right panel; bottom plot of Figure \ref{f_vvmax}).
The bias in absolute magnitude is now 0.096 mag and the detection
efficiency is reduced by 88.6\% with respect to Case I.

These exercises show that the mean absolute magnitude of the detected
sample (the zero-point of the Hubble diagram) could be substantially biased
because the detected sample contains a relatively larger fraction of
luminous events (with broader light curves) than that of the parent
population.  However, the existence of the M/$\Delta$m$_{15}$ relation
allows us to correct for the luminosity excess of these SNe.  In the
limiting cases considered above, where the scatter about this relation
is zero, there is a one-to-one mapping between absolute magnitude and
decline rate, so that the mean {\em corrected} absolute magnitude has
identically zero bias. In other words, both the slope and the zero-point of
the parent function remain unaffected by the selection effects.

In reality, of course, there is significant intrinsic scatter in the peak
magniudes and we expect a bias in the mean {\it corrected} absolute
magnitude because the selection effects will tend to pick the brightest
events from the corrected peak luminosity distribution.  We now estimate
the magnitude of this bias under the assumptions that the parent
M/$\Delta$m$_{15}$ function has a dispersion due to the peculiar motion of
the host galaxies ($\sigma_{V}$=600 km sec$^{-1}$) and an additional
scatter in absolute magnitude independent of the SN lightcurve shape
($\sigma_{M}$=0.15).  To make the simulation somewhat more realistic, we
also assume that the limiting magnitude for SN detection has a fluctuation
of $\sigma_{Blim}$=0.3.

Figure \ref{f_bias} (top panel) presents the results of this simulation in
the form of absolute magnitudes of the detected SNe plotted versus
$\Delta$m$_{15}$(B).  We note first that the density of points in this
diagram is significantly lower for dimmer events since, as expected, these
SNe are more difficult to discover. The dashed line is the least-squares
fit (taking into account errors in absolute magnitudes) to the detected SNe
\footnote[3] {An examination of the residuals of the 1,000 fake SNe about
the least-squares fit shows, in fact, that they have a Gaussian
distribution.}.  This should be compared with the solid line (almost
indistinguishable) that corresponds to the parent relation (with
coefficients a$_{0}$=-19.258 and b$_{0}$=0.784.).  The fit to these 1,000
SNe gives a zero-point of a=-19.30$\pm$0.01 and a slope of b=0.76$\pm$0.02.
Selection effects can thus introduce a systematic bias of 0.04 mag in the
zero-point of the observed relation, less than half the 0.10 mag bias
in the uncorrected absolute magnitudes. This is because the detected sample
suffers only from the bias left {\it after} correcting for the decline
rate.  When the scatter of this relation is increased to
$\sigma_{M}$=0.50 (a value greatly in excess of that observed), we obtain
a=-19.59$\pm$0.02 and b=0.78$\pm$0.06 (second panel of Figure
\ref{f_bias}), and the much larger bias of 0.33 mag in the zero-point.  It
is interesting to note that the slope of the parent function is almost
unaffected even by such a large scatter about the relation.  The magnitude
of the bias depends only on the {\em dispersion} of the absolute magnitudes
and not on their {\em values}. Because we assume the same dispersion for
all lightcurve widths, the magnitude of the bias is the same for all
widths.

In summary, this test shows that a dispersion about the parent
M/$\Delta$m$_{15}$ function results only in a bias of the zero-point of the
observed relation; the slope of the observed relation remains
unaffected. More generally, the shape of the observed relation will be
the same as that of the parent function, no matter what the shape of the
parent function, {\em under the assumption that the scatter around the
function is independent of lightcurve width}. To illustrate this, we show
the results of a simulation with a quadratic function for the
M/$\Delta$m$_{15}$ relation, with coefficients a$_{0}$=-19.258,
b$_{0}$=0.457, and c$_{0}$=1.293 corresponding to the preliminary
parameters found by \cite{phillips98}. To enhance the biases from selection
effects we once again adopt the excessively large scatter of
$\sigma_{M}$=0.50. A quadratic fit to the detected SNe yields
a=-19.59$\pm$0.02, b=0.49$\pm$0.12, and c=1.15$\pm$0.23.  As anticipated,
the linear and quadratic coefficients remain unchanged (within the error
bars) and the bias of 0.33 mag in the zero-point is identical to that of
the previous case.  The third panel of Figure \ref{f_bias} shows very
clearly that the shape of the parent function is unchanged.

A different result obtains when the dispersion about the M/$\Delta$m$_{15}$
function is correlated with lightcurve width. If we assume that the scatter
ranges from $\sigma_{M}$=0.15 at the smallest decline rate to
$\sigma_{M}$=0.50 at the largest, the magnitude of the bias increases with
decline rate, resulting in a change in the slope of the M/$\Delta$m$_{15}$
relation as seen in the bottom panel of Figure \ref{f_bias}. The fit to the
detected events yields a=-19.34$\pm$0.01 and b=0.51$\pm$0.04. Both the
zero-point and the shape of the parent function might thus be substantially
biased under these circumstances.

A related question is whether the observed zero-point and slope could be
substantially in error purely by chance due to the small number of events
(27) used in their determination.  To answer this question we estimate the
uncertainties in the a and b coefficients by running a large number of
simulations for the case $\sigma_{M}$=0.15, always fixing the number of
detections to 27.  From a run of 200 simulations we get a=-19.29$\pm$0.04
and b=0.77$\pm$0.14 where the quoted uncertainties (standard deviations)
reflect the fluctuations of these parameters due to statistical
errors. From a single experiment with 27 detected SNe the observed
zero-point and slope are expected to have random errors of only 0.04 mag
and $\sim$20\%, respectively; these are almost identical to those quoted in
Paper V.

We can assess the reality of a M/$\Delta$m$_{15}$ relation by a similar
experiment, asking ``What is the probalibity that the CT sample could yield,
by chance, the observed slope given the hypothesis of a null slope?''  From
a run of 200 simulations and the initial assumption that b$_{0}$=0 and
$\sigma_{M}$=0.15 (again fixing the number of detections to 27), we obtain a
nearly Gaussian distribution of slopes with b=-0.009$\pm$0.12.  The value
of of b=0.784$\pm$0.182 derived in Paper V thus lies 3.6 $\sigma$ away from
the assumption of a null slope and the data of the CT survey show this
hypothesis can be rejected with 99.97\% confidence.

In Paper V concern was expressed about the disagreement in the slopes of
the M/$\Delta$m$_{15}$ relation inferred from the CT SNe and from Phillips'
original sample \footnote[4]{This sample was selected by having 1) precise
optical photometry, 2) well-sampled light curves, and 3) accurate relative
SBF and PNLF distances (at the distances of the Virgo and Fornax clusters
or closer).}.  The disagreement can be seen very clearly in Figure 2 of
Paper V and, also, by performing a linear fit to the data (given in Table 2
of Paper V) of the nearby sample of eight SNe with
0.87$\leq\Delta$m$_{15}$(B)$\leq$1.73 (the same range of decline rates used
to determine equation (\ref{eqn1})), which yields a=-19.18$\pm$0.08 and
b=1.89$\pm$0.34. The likelihood that this result is consistent with the CT
slope can be evaluated from Monte Carlo simulations.  Although the SNe of
Phillips' sample are not part of a systematic survey like the CT search,
for the sake of simplicity we will assume in the following test the same
selection criteria.  From a run of 200 simulations (fixing the number of
observations to that of the nearby sample) and the initial assumptions that
b$_{0}$=0.784 and $\sigma_{M}$=0.15 (the CT parameters), we obtain
b=0.764$\pm$0.31. The slope of the nearby sample thus differs from the CT
slope by 2.4 $\sigma$, i.e, the likelihood of obtaining the slope of the
nearby sample by pure chance from a parent population like the CT sample is
less than 2\%.

As a possible explanation for this evident discrepancy, in Paper V it was
speculated that this could be due to biases in the CT search. As discussed
above, one possibility of reconciling the shallower slope from the CT
sample with that obtained from the nearby sample is that the dispersion
about the parent M/$\Delta$m$_{15}$ relation increases toward large decline
rates.  To examine this possibility in more detail we use the zero-point
and the slope obtained from the nearby sample as parent coefficients to ask
how much scatter in the parent function is required in order to get the
slope of the CT sample.  We find that to match this slope one must increase
the scatter about the relation from 0.15 mag for the slowest-declining
events to 0.95 mag for events with $\Delta$m$_{15}$(B)$\sim$1.7. An
unsatisfactory consequence of this large dispersion in the parent
distribution, however, is that the sample of detected events also displays
an exceedingly large dispersion at low decline rates, a feature which
contrasts with the relatively small and constant scatter observed about the
CT M/$\Delta$m$_{15}$ relation. Thus, a strongly-increasing scatter
around the actual relation in the range
0.87$\leq\Delta$m$_{15}$(B)$\leq$1.73 does not appear to be a viable
explanation.  As was mentioned in Paper V, another possible source of
discrepancy between the two data samples is the uncertainty in the SBF/PNLF
distances employed in the nearby sample.  It is also possible that
corrections for extinction from dust could lead to a better agreement
between the two samples.

In summary, the results of this section and the small scatter observed in
the CT M/$\Delta$m$_{15}$ relation in the range
0.87$\leq\Delta$m$_{15}$(B)$\leq$1.70 suggest that the bias in the observed
slope is not large and the observed zero-point is expected to be biased by
no more than $\sim$0.04 mag.  Additional tests performed with different
values of M$^{\rm B}_{1.1}$ show that these findings are insensitive to the
choice of the absolute magnitude for the SNe. This result is further
insensitive to the shape of the SN~Ia luminosity function. This supports
our initial assumptions regarding equation (\ref{eqn1}) in the range
0.87$\leq\Delta$m$_{15}$(B)$\leq$1.70.  As we mentioned above, however, it
is troubling that the relation overestimates the absolute magnitude of SN
1992K ($\Delta$m$_{15}$(B)=1.93) by 2$\sigma$.  Further evidence for a
departure of fast-declining events from the linear approximation is
provided by the SN 92K-like event in the nearby sample (SN 1991bg) and by
the recently discovered SN 1997cn (\cite{turatto98}). In Section 5 we
consider a M/$\Delta$m$_{15}$ relation that accounts for these facts.

\subsection{The luminosity function of SNe~Ia}

In this section we study possible biases in determining the luminosity
function of SNe~Ia.  Figure \ref{f_eff} (left panel) shows with solid
circles the resulting decline rate distribution of the 50,000 SNe detected
in a simulation performed with the standard set of parameters [M$^{\rm
B}_{1.1}$=-19.258 (H$_{0}$=65), B$_{lim}$=18.3, $\alpha$=0.15
kpc$^{-1}$]. Also shown in this figure is the histogram of the CT
sample. The comparison reveals a relatively good agreement between the
simulation and the data (with a KS probability of 0.51), which suggests
that the initial assumption of a flat decline rate distribution in the
simulation is not a bad guess. Since, by construction, the parent
population in the simulation has equal numbers of SNe per bin in decline
rate, the solid circles represent a measure of the detection efficiency.
This curve reveals an increasing incompleteness toward higher decline rates
which is a clear signature of the M/$\Delta$m$_{15}$ relation (equation
(\ref{eqn1})); events with faster decline rates are intrinsically fainter
and more transient and, hence, more difficult to detect.  The shape of this
function proves to be independent of the adopted value for M$^{\rm
B}_{1.1}$ and we can employ it in principle to correct the observed
histogram for incompleteness. Figure \ref{f_imf} shows the ``corrected''
luminosity function obtained by dividing the observed histogram of decline
rates by the efficiency curve (properly normalized to preserve the total
number of objects) and transforming the decline rates of the abscissa to
absolute magnitudes through equation (\ref{eqn1}).  As anticipated, the
``corrected'' distribution is remarkably flat over the whole range of
decline rates with, perhaps, an increase in frequency for the intrinsically
faint events as indicated by the straight line (an approximate visual fit
to the ``corrected'' distribution).  It is important to note that this
result is very sensitive to the M/$\Delta$m$_{15}$ relation adopted and,
hence, is somewhat uncertain.  For example, if the actual slope of this
relation was steeper than that assumed in the simulation, the
efficiency curve would become steeper and the ``corrected'' distribution
would display an even larger number of faint SNe.

Another important caution regarding the ``corrected'' distribution is the
possible non-linearity of the M/$\Delta$m$_{15}$ relation toward high
decline rates. With M$^{B}_{MAX}\sim$-17.7+5log(H$_{0}$/65), SN 1992K is
$\sim$0.9 mag fainter than the value obtained from the adopted relation
(equation (1)) for a decline rate of $\Delta$m$_{15}$(B)=1.93.
Consequently, SN 1992K-like events are not produced in the simulation.
This can be seen in the right panel of Figure \ref{f_eff} which compares
the absolute magnitude distribution of the 50,000 simulated SNe with that
of the CT sample. Although there is overall agreement between the
simulation and the data, the lack of simulated events like the subluminous
SN 1992K is evident. To account for these subluminous SNe it would be
necessary to modify the parent M/$\Delta$m$_{15}$ function by adopting a
steeper slope toward the highest decline rates.  In such a case, the actual
efficiency for detection of the subluminous events in the simulation would
be much lower than what we estimate here and, hence, the actual frequency
of the fast-declining SNe would be substantially higher than that shown in
the ``corrected'' histogram of Figure \ref{f_imf}.  In summary, the
``corrected'' decline-rate distribution should be much more reliable in the
range 0.87$\leq\Delta$m$_{15}$(B)$\leq$1.7 where the M/$\Delta$m$_{15}$
relation is well represented by equation (\ref{eqn1}) (see Paper V).

\subsection{The galaxy type-redshift relation}

One of intriguing features of the CT sample is that the ratio of elliptical
to spiral hosts (hereafter E/S) increases dramatically with redshift (see
Figure 4 of Paper VI). Clearly a selection bias must be at work here
because it is hard to imagine that evolutionary effects could be this
important at redshifts below than 0.1.  Within the framework of our simple
model the only agent that can explain this feature is the Shaw effect.

At zero redshift, the minimum angular separation maps into a vanishingly
small region of the host so that in principle nearly all SNe can be
discovered in nearby galaxies. At larger redshifts, the same separation
maps into a larger radius of the parent galaxy so that an increasingly
larger volume of the host remains unsurveyed; only a fraction of the SNe
are potentially discoverable. The E/S ratio can change with redshift since
the mass of this ``exclusion'' region varies with radius differently for
different morphological types.  This variation can be determined
analytically, assuming that the SN radial distribution follows the light
distribution, by integrating the light profile from an inner radius out to
infinity.  The inner radius  defines the outer bound of the ``exclusion'' region
and is linearly related to the distance (or
redshift) of each host.  The top panel of Figure \ref{f_morph} illustrates
this effect.  The solid lines show the fraction of the total host light
that is sampled in the SN survey as a function of redshift (H$_{0}$=65) for
an elliptical with r$_{eff}$=6 kpc and a spiral with $\alpha$=0.15
kpc$^{-1}$ (r$_0$=6.7 kpc) (for the latter this curve is calculated from
the {\it projected} light distribution on the plane of the sky, averaged
over all disk inclinations). Since the fraction of sampled light for
spirals differs at any given redshift from that of ellipticals, the E/S
ratio varies with redshift. The dashed line shows the E/S ratio,
arbitrarily normalized to unity at zero redshift.  The E/S ratio increases
strongly at high redshifts as the light from an r$^{1/4}$ law declines more
slowly with radius than that from an exponential disk. This effect can be
seen in the simulations as well. The bottom
panel of Figure \ref{f_morph} shows the comparison between the
analytic E/S curve (solid line) and simulations (solid points) performed
with the same parameters (r$_{eff}$=6 kpc, r$_0$=6.7 kpc) based on 10$^6$
detected events.

Although the magnitude of this effect in our simple model is not sufficient
to account for the CT E/S ratio (which increases from 0.14$\pm$0.15 in the
bin v$<$10,000 km sec$^{-1}$ to 1.33$\pm$0.72 in the bin v$>$10,000 km
sec$^{-1}$), the E/S ratio is very sensitive to the values of r$_{eff}$ and
r$_0$. The dashed line in the figure is determined from r$_{eff}$=6 kpc and
r$_0$=4 kpc.  Note the sudden rise of the E/S ratio at lower
redshifts. This exercise merely demonstrates that while there may be other
selection effects responsible for this feature in the CT sample, the Shaw
effect can modify the E/S ratio in the same direction as that observed.

\section{A More Speculative Model}

Figure \ref{f_simpdmdb} shows the individual apparent magnitudes, absolute
magnitudes, and decline rates of a random subsample of 1,000 SNe detected
in the simulation performed with the simple model described above and the
standard set of parameters, as functions of recession velocity. The large
circles show the CT SNe.  While the match to the observations is
encouraging, a few deficiencies are obvious. In the middle panel it is
clear that the simulation lacks faint events like SN 1992K at
M$^{B}_{MAX}\sim$-17.7+5log(H$_{0}$/65).  As pointed out in Paper VI, it is
also clear that the CT sample lacks bright events in the bin of distant SNe
(log(v)$\geq$4) (an anti-Malmquist bias?).

Another problem is posed by the two most distant SNe in the observed sample
which appear to have been detected well beyond the nominal detection limit
of the simulation (corresponding to B$_{lim}$=18.3) represented by the
dashed diagonal line in the middle panel. Although the detection of events
beyond this limit can occur in the simulation (since we allow for
fluctuations of the limiting magnitude), statistically this is insufficient
to explain the presence of these two outliers.

In Figure \ref{f_simpshaw} we show the individual apparent magnitudes,
absolute magnitudes, and decline rates of the same subsample of 1,000 SNe,
as functions of the projected distance between the SN and the center of the
host galaxy. The diagonal line in the top panel represents the boundary of
the central 4 arcsec where no SNe were discovered.  Not surprisingly, the
simulations match this feature. In the other panels the agreement is
generally good, although it is again evident that the simulations lack dim
SNe (central panel).  

Having estimated biases in some of the observed quantities, we can use this
information to review our assumptions and attempt to resolve some of
these problems. We introduce the following modifications to our initial
assumptions about the parent M/$\Delta$m$_{15}$ relation and the SN
luminosity function.

\noindent $\bullet$ The lack of intrinsically faint SNe in the simulation
suggests we modify the assumed M/$\Delta$m$_{15}$ function. Because the
tests performed in Section 4.1 suggest that the observed relation is not
severely biased by the selection effects, we will still adopt the relation
given by equation (\ref{eqn1}) and the coefficients a$_{0}$=-19.258 and
b$_{0}$=0.784 for the range 0.87$\leq\Delta$m$_{15}$(B)$\leq$1.7.  For the
range 1.7$\leq\Delta$m$_{15}$(B)$\leq$1.93, however, we will assume a
steeper slope which describes the absolute magnitude of SN 1992K and
ensures the continuity of the parent function.  With these constraints the
zero-point and slope of this function in the range
1.7$\leq\Delta$m$_{15}$(B)$\leq$1.93 are -21.573 and 4.642, respectively.
Note that this ``two-slope'' model is not much different than the quadratic
form recently used by \cite{phillips98} for the M/$\Delta$m$_{15}$ relation
after applying corrections for host galaxy extinction.  Although it would
be preferable to employ corrected absolute magnitudes in our simulations,
these data are not yet available.

\noindent $\bullet$ Our examination of bias in the observed luminosity
function suggests that the frequency of the parent SN population might
increase toward fast decline rates. In the following we will assume a
parent luminosity function with the linear form shown in Figure
\ref{f_imf}, i.e,
\begin{equation}
N = k \times\Delta m_{15}({\rm B}),
\end{equation}
where k is a normalization. With this approximation, the fastest-declining
observed events occur with approximately twice the frequency of the
slowest-declining.  This is reasonable as we found above that the corrected
frequency of the fastest-declining events would have been higher than that
shown in Figure \ref{f_imf} had we employed a steeper M/$\Delta$m$_{15}$
relation for these events.

Using these revisions in our description of the parent population, we
perform a parameter search in the range -19.558$\leq$M$^{\rm
B}_{1.1}\leq$-18.958, 17.3$\leq$B$_{lim}\leq$18.7, 0.05$\leq\alpha\leq$0.40
kpc$^{-1}$. We have also relaxed the constraint of a fixed $\sigma_{Blim}$
and vary this parameter in the range 0.3-1.2 mag, allowing detection of
events well beyond the nominal detection limit of the simulation. We then
look for the set of parameters that {\it simultaneously} yields the highest
KS probabilities for the apparent magnitude, recession velocity, angular
separation, and projected radial distance between the SN and the host
galaxy nucleus. Note that in this case the solutions are much more severely
constrained than for the simple model (where we searched for parameters
which fit only the recession velocity and the angular separation
distributions).

With these requirements we find that the best solution in the
($\alpha$,B$_{lim}$,$\sigma_{Blim}$) parameter space changes little within
the range of absolute magnitudes considered. For the specific case of
M$^{\rm B}_{1.1}$=-19.258 (H$_{0}$=65), the best solution is obtained for
$\alpha$=0.175 kpc$^{-1}$, B$_{lim}$=17.9, and $\sigma_{Blim}$=0.7.  This
last value is significantly larger than our initial assumption of
$\sigma_{Blim}$=0.3.  Figure \ref{f_bestfour} shows the recession velocity,
apparent magnitude, angular separation, and radial distance distributions
of 50,000 detected SNe in a simulation performed with these parameters
along with the observed supernov\ae. The match to the observations is
remarkably good with KS probabilities 0.99, 0.76, 0.96, and 0.46,
respectively, and evidently better than that shown in Figure
\ref{f_simphist} for the simpler model.  Figure \ref{f_besttwo} shows the
decline rate and absolute magnitude distributions, with KS probabilities of
0.81 and 0.86, respectively. Again, the agreement with the observations is
remarkably good and certainly much better than that obtained with the
simple model (cf. Figure \ref{f_eff}).  One interesting feature is the long
tail of the absolute magnitude distribution toward intrinsically faint SNe,
a consequence of the steeper slope adopted for the M/$\Delta$m$_{15}$
relation. The decline rate distribution of the simulation has a shape that
closely follows the observed histogram, suggesting that relaxing the
assumption of a flat luminosity function is a real improvement.

While these modifications to our simple model improve agreement with
observation, they are admittedly somewhat {\em ad hoc} and one might
question whether some other change would prove similarly attractive.
While there is no way to systematically examine the space of
possible models, as an example we examine the consequences of another
possibility.

We adopt instead the hypothesis that a flat luminosity function should be
extended to fainter absolute magnitudes (ELF).  Events with with
$\Delta$m$_{15}$(B)$>$1.93 have never been observed, and we must guess at
the lightcurve shape of such events. We have extrapolated the set of six
known templates to $\Delta$m$_{15}$(B)=3 and extended the ``two-slope''
M/$\Delta$m$_{15}$ relation to higher decline rates.  Adopting
M$^{\rm B}_{1.1}$=-19.258, an examination of the parameter space yields
a best solution for $\alpha$=0.2 kpc$^{-1}$, B$_{lim}$=17.7, and
$\sigma_{Blim}$=0.9, results very similar to the previous case.
For these parameters the KS probabilities for recession
velocity, apparent magnitude, angular separation, and radial distance
distributions of 50,000 detected SNe are 0.97, 0.88, 0.58, and 0.61,
respectively, and are comparable to those obtained for the
non-flat luminosity function.

The decline rate and absolute magnitude distributions, on the other hand,
give KS probabilities of 0.54 and 0.42, significantly lower than those
obtained with the non-flat luminosity function (0.81 and 0.86).  This drop
in the KS probabilities characterizes the whole parameter space
explored. For 600 parameter combinations, the average KS
probability for $\Delta$m$_{15}$(B) is 0.77 in the non-flat case and 0.54
in the ELF case. For absolute magnitudes, the probabilities are
0.86 and 0.41, respectively. This exercise shows that the Monte Carlo
technique and the KS test have some power to distinguish between different
models despite the limitations imposed by the low number of detected
events. In particular, we can safely conclude that an ELF is less
consistent with the data than the non-flat luminosity function.

Returning to the favored model, Figure \ref{f_bestdmdb} shows the
individual apparent magnitudes, absolute magnitudes, and decline rates of a
random subsample of 1,000 SNe detected in the simulation performed with the
best parameters as functions of recession velocity. The large circles show
the CT SNe which are in very good agreement with the simulations. A
noticeable difference with the simple model (Figure \ref{f_simpdmdb}) is
the large scatter displayed by the absolute magnitudes in the nearby bin
and the presence of SN 1992K-like objects as in the CT sample.  This larger
scatter is due in part to the new population of intrinsically faint SNe
which cannot be detected at very large distances and do not appear in the
distant bin (log(v)$\geq$4). Another effect that contributes to the larger
dispersion in the nearby bin is the peculiar motion of the host galaxies
(600 km sec$^{-1}$) which constitutes a larger fraction of the recession
velocity of the nearby events. In the distant bin, on the other hand, the
discovered SNe appear to have a lower dispersion than the simulated
sample. Although these objects cluster precisely in the region of the
diagram with the highest density of simulated SNe, it is possible that the
lack of intrinsically bright events in the distant bin is a signature of a
significant difference between the simulations and the data.

Another improvement with respect to Figure \ref{f_simpdmdb} is that the two
most distant CT SNe appear to be much closer to the bulk of synthetic SNe,
a result of the large fluctuation of 0.7 mag in the limiting magnitude of
the simulation.  Even better agreement would have been obtained if these
two distant SNe had brighter intrinsic luminosities (or, equivalently,
smaller decline rates), since these are the events which are most likely
found near the detection limit of the simulation. In any case, it is hard
to ascertain with such few objects whether or not this is a significant
difference. The distribution of decline rates (bottom panel) is similar to
that of absolute magnitudes as a consequence of the M/$\Delta$m$_{15}$
relation. The sharp cutoffs at small and large $\Delta$m$_{15}$(B) are
there by construction, as explained in section 3.1.

Figure \ref{f_bestshaw} shows the individual apparent magnitudes, absolute
magnitudes, and decline rates of the same subsample of 1,000 SNe, as
functions of the projected distance between the SN and the host galaxy
nucleus. A comparison with Figure \ref{f_simpshaw} shows a noticeable
improvement over the simple model.  Motivated by the claim by \cite{wang97}
that the dispersion in absolute magnitude among the CT SNe decreases with
galactocentric distance, we ask whether this feature is present in the
sample of simulated SNe as a consequence of the Shaw effect.  Figure
\ref{f_rdist} shows the absolute magnitude distributions from the sample of
50,000 synthetic SNe separated into two bins: events with projected radial
distances larger than 20 kpc (solid line) and those with distances smaller
than 10 kpc (dashed line). Despite a small (but real) difference due to the
Shaw effect, the scatter in absolute magnitude does not
change dramatically from one bin to the other. This may mean that the model
is unable to reproduce this feature, but it is equally likely that the
effect seen by \cite{wang97} is merely a statistical fluctuation.  In fact,
most of the dispersion in absolute magnitude among the CT events is caused
by one object, SN 1992K.  It remains to be seen if this correlation will
persist as the diagram is populated with more SNe.

Finally, Figure \ref{f_timedisc} (top panel) shows the age at discovery of
the 50,000 SNe plotted as a function of recession velocity. This
illustrates very clearly the effect that distant events can be found only
during a short period of time around maximum light. For comparison, the
bottom panel of Figure \ref{f_timedisc} shows the corresponding plot for
the CT SNe, revealing a satisfactory agreement with the simulation.

\section{The SNe~Ia Rate}

The actual rate of SNe~Ia (in units of SNe/volume/time), $\dot{n}$, can in
principle be calculated from the number of discovered SNe in the CT survey,
N$_{dis}$, and the detection efficiency of the search (in units of volume
$\times$ time), {\it eff}, which we define as
\begin{equation}
eff = \frac{N_{dis}}{\dot{n}}.
\label{eqn4}
\end{equation}

In the following we assume that the actual efficiency of the search
corresponds to that of the simulation performed with the parameters that
best fit the CT data.  The efficiency of the simulation can be easily
calculated from the number of detected SNe, N$_{det}$, and the input rate
which can be computed from the number of generated SNe, N$_{gen}$, the
volume and the period of time over which the events are distributed in the
simulation.  Since we generate SNe in a spherical volume of radius R (in
units of Mpc) over a period of 4 years, the efficiency of detection (in
units of Mpc$^{3}$ $\times$ century) appropriate for the CT survey is,
\begin{equation}
eff = \frac{N_{det}}{N_{gen}} \times \frac{4\pi R^{3}}{3} \times \frac{4}{100} \times
\frac{60 \times 26.129}{41,253}.
\label{eqn5}
\end{equation}
The 4/100 factor converts the rate from units of 4 years to 100 years.  The
last factor in this equation corresponds to the fraction of the celestial
sphere (41,253 square degrees) surveyed by the CT project (a total of 60
fields each providing a sky coverage of 26.129 square degrees), which is
appropriate since the simulations do not constrain the detection of SNe by
their location in the sky. From equations (\ref{eqn4}) and (\ref{eqn5}), it
is possible to solve directly for $\dot{n}$, the actual rate of SNe~Ia in
units of SNe Mpc$^{-3}$ century$^{-1}$. However, SN rates are usually
expressed in units of ``SNu'': SNe per 10$^{10}$ L$_{\odot}$ per century
(where L$_{\odot}$ corresponds to the B solar luminosity). We denote the
rate in units of SNu by $\nu$ (\cite{capellaro97}); its numerical value can
be obtained by dividing $\dot{n}$ by L$_{B}$, the blue luminosity density
of galaxies (in units of 10$^{10}$ L$_{\odot}$ Mpc$^{-3}$) of the local
Universe.  Marzke (private communication) has kindly provided us with a
recent determination of the blue luminosity density of
\begin{equation}
L_{B} = 1.65 (\pm 0.42) \times 10^{-2} h \left[10^{10} L_{\odot}
Mpc^{-3}\right],
\label{eqn6}
\end{equation}
where h=H$_{0}$/100. This value of L$_{B}$ has been calculated from
asymptotic magnitudes B$_{T}$ which are defined to account for the whole
integrated light of the galaxies of his sample (see \cite{devaucou91} for
the definition of B$_{T}$).  Combining equations (\ref{eqn4}),(\ref{eqn5}),
and (\ref{eqn6}), the SN~Ia rate in SNu is,
\begin{equation}
\nu = 31 \times \frac{N_{gen}}{N_{det}} \times \frac{3}{4\pi R^{3}} \times \frac{100}{4} \times \frac{41,253}{60 \times 26.129} \times \frac{1}{1.65 \times 10^{-2} \times h}
\label{eqn7}
\end{equation}
where we have replaced N$_{dis}$ by the total number (31) of SNe~Ia events
discovered in the course of the CT survey.

Table 4 gives the parameters $\alpha$, B$_{lim}$, and $\sigma_{Blim}$ that
best fit the CT observations for M$^{\rm B}_{1.1}$=-19.258, as described in Section
5, along with the values of N$_{gen}$, N$_{det}$, the radius R of the
simulation run, and the corresponding rate computed with equation
(\ref{eqn7}). A large value of N$_{det}$ was chosen to minimize the
statistical error in the determination of the detection efficiency.  The
uncertainty quoted for the calculated rates is the sum in quadrature of the
following errors, which we assume to be uncorrelated:

\noindent $\bullet$ As in all counting experiments, we expect statistical
fluctuations in N$_{dis}$ which we estimate by running many simulations,
always fixing N$_{det}$ to 31 and computing the standard deviation of the
rate.  The error due to the statistical fluctuations amounts to $\sim$16\%
(which proves to be comparable to $\sim\surd$N/N$\sim$18\%).

\noindent $\bullet$ The uncertainty in $\alpha$, which we assume to be
$\pm$0.05 kpc$^{-1}$.  This error amounts to $\sim$13\%.

\noindent $\bullet$ The uncertainty in B$_{lim}$, which we assume to be
equal to $\sigma_{Blim}$ ($\pm$0.7 mag).  The error bar in the resulting
SN rate is not symmetric and amounts to $\sim$50-150\%.

\noindent $\bullet$ The uncertainty in the luminosity density, which
amounts to 25\% (equation (\ref{eqn6})) in the SN rate.

\noindent Clearly (and somewhat remarkably), the largest source of
uncertainty in the calculated rates is the lack of a precise value for
B$_{lim}$ and not the small number of SNe found in the survey.

Recently, \cite{capellaro97} employed a sample of 7,773 galaxies from the
Third Reference Catalogue of Bright Galaxies (\cite{devaucou91}, hereafter
RC3) and the combined sample of SNe from five searches to compute rates of
SNe using the ``control time method''.  Although the CT SNe were included
in this calculation, these objects constitute only 10\% of the total sample
so that the two estimates are quite independent of each other. Also, while
we have searched a fixed volume of space for supernov\ae\ in a systematic
manner, \cite{capellaro97} use a fixed set of galaxies and take the results
of a heterogeneous set of searches.  \cite{capellaro97} report an average
rate of SNe~Ia of $\nu$=0.20 ($\pm$ 0.07) (H$_{0}$/75)$^{2}$ SNu. Since the
normalization to galaxy luminosity of their work is based on the
extinction-corrected B$_{T}^{0}$ magnitudes of the RC3 catalogue, their
value cannot be directly compared to our rate which is based on
uncorrected B$_{T}$ magnitudes. To allow a proper comparison between the
two calculations, Capellaro (private communication) has kindly provided an
alternative rate for SNe~Ia of 0.22 ($\pm$ 0.07) (H$_{0}$/75)$^{2}$ SNu,
duly normalized to the galaxy luminosity {\it uncorrected for internal
extinction}. Given the uncertainties, Capellaro's rate of 0.161$\pm$0.051
(for H$_{0}$=65) agrees surprisingly well with our estimate of 0.21$^{\rm
+0.30}_{-0.13}$.

\section{Discussion and Conclusions}

We have shown that a simple model of the selection effects of the CT
photographic survey, coupled with the assumptions of a linear
M/$\Delta$m$_{15}$ relation and a flat luminosity function for the parent
population of SNe~Ia accounts for the basic properties of the observed SN
sample. The model accounts for biases due to the flux limited nature of the
survey, the different light curve morphologies displayed by different
SNe~Ia, and the difficulty of finding SNe projected near the central parts
of the host galaxies (\cite{shaw79}).

Using the simple model we have estimated the bias in the zero-point and the
slope of the M/$\Delta$m$_{15}$ relation. For an assumed scatter of 0.15
mag about this relation, the selection effects of the search (as modeled)
decrease the zero-point by 0.04 mag (and, hence, decrease the actual value
of H$_{0}$ by 2\%).  The estimated bias would thus have a negligible effect
in the determination of cosmological parameters from high redshift SNe.  If
the M/$\Delta$m$_{15}$ relation is ignored, the bias in the zero-point of
the Hubble diagram proves to be at least 0.10 mag. The slope of the
relation, however, does not appear to be substantially affected
by selection effects in the range 0.87$\leq\Delta$m$_{15}$(B)$\leq$1.7. It
appears, then, that the shape and zero-point of the M/$\Delta$m$_{15}$
relation determined from the CT SN sample are quite reliable.

We have estimated the degree of incompleteness of the survey as a function
of decline rate and used this efficiency curve to produce a corrected
luminosity function for SNe~Ia in which the frequency of SNe increases with
decline rate. This result is somewhat uncertain as it is very sensitive to
the adopted M/$\Delta$m$_{15}$ function.

With the simple model and radial light profiles for the host galaxies
assuming r$^{1/4}$ and exponential laws for elliptical and spiral galaxies,
respectively, we have shown that the Shaw effect will affect the observed
ratio of elliptical to spiral hosts as a function of redshift, an effect
present in the CT sample.

Based on these results we have proposed a new model for the parent
population of SNe which better accounts for the properties of the observed
SN sample. This assumes a non-linear M/$\Delta$m$_{15}$ relation and a
skewed luminosity function. Not surprisingly, with the larger number of
degrees of freedom this model allows, the simulations provide a much better
match to the observations than that obtained with the simple model.

Since our goal was to propose a model of the selection effects of the CT
survey that entails the least possible number of assumptions and
parameters, we have made no consideration of an additional potential bias
due to extinction by dust in the host galaxy (the inclusion of dust in the
model would require at least three additional parameters to describe the
geometry and magnitude of the extinction).  Although the success of this
model may provide some indication that these additional selection biases
are small, at least in surveys performed with photographic techniques, it
is fair to ask how important dust extinction may be among SNe~Ia.

\cite{capellaro97} find that dust absorption in the host galaxy causes
SNe~Ia to be detected 1.8$\pm$0.5 times more frequently in spirals with low
inclinations. \cite{vdb90} find that SNe~Ia do not exhibit such an
inclination effect.  For the specific sample of CT SNe, the spectroscopic
data of these objects revealed only moderate amounts of extinction by dust
(Paper VI), in agreement with the conclusion of \cite{vdb90}. This is
confirmed by the color excesses calculated from the observed
(B$_{MAX}$-V$_{MAX}$) colors of the CT sample and a preliminary
(B$_{MAX}$-V$_{MAX}$) intrinsic color versus decline rate relation kindly
provided to us before publication (\cite{phillips98}). Definitive
reddenings and corrected absolute magnitudes will be soon available.
Figure \ref{f_dust} shows the color excesses of the CT SNe~Ia due to
extinction in the host galaxy as a function of the projected radial
distance.  As expected, the extinction increases toward the center of the
spiral galaxies (solid dots) where the dust density may be expected to be
higher. The effect of dust on the ensemble of CT SNe~Ia, however, is not
very large (the average E(B-V) is less than 0.1). This is consistent with
the success of the model presented in this paper.  In a future paper we
will include a reddening model in the simulations, once these
reddening-corrected absolute magnitudes for the CT SNe become available.

In a recent paper \cite{hatano98} proposed a model of the spatial
distribution of dust and SN progenitors which produces a large dispersion
in the absolute magnitudes of the SNe~Ia projected on the nuclear regions
of the host galaxies (see their Figure 3). An obvious problem with this
model is that this scatter proves to be three times larger than the
dispersion implied by the color excesses of the CT SNe (shown in Figure
\ref{f_dust}).  An additional problem with this model is their choice of
the luminosity function for SNe~Ia, M$^{B}$$_{MAX}$=-19.2$\pm$0.2, which
does not account for the population of intrinsically faint SNe.  The
inclusion of faint SNe would increase the scatter in luminosity still
further, making the comparison with observation even worse.  It appears
likely that while dust plays a role in explaining some of the luminosity
dispersion of the SNe projected near the central parts of the host
galaxies, it is not as important as \cite{hatano98} have claimed.

We have assumed that SNe~Ia occur in idealized host galaxies consisting of
smooth exponential disks, with the radial distribution of SNe following
that of the host galaxy light. Since the Shaw effect can modify the
observed radial distribution of SNe, we included this selection effect in
our simulation. This showed that the effect is too small to account for the
observed increase in scatter in absolute magnitudes at small projected
radial distances, first noticed by \cite{wang97} among the CT SNe. A
probable cause for this disagreement is simply small number statistics.
Most of the dispersion among the CT SNe is caused by one object, SN 1992K.
The most likely alternative is dust absorption.  From Figure \ref{f_dust}
it is clear that the highest extinction occurs at the smallest
galactocentric distances, so some of the dispersion could be due to
uncorrected dust absorption. Part of the observed fluctuation is due to
over-luminous events, however, which cannot be the result of
obscuration. We have postponed any further consideration of the effects of
dust until corrected magnitudes become available (\cite{phillips98}).  We
note as well that since elliptical galaxies dominate the hosts at large
separations (due to the extended distribution of light from elliptical
galaxies and the Shaw effect, as suggested in section 4.3), the observed
decrease in the absolute magnitude scatter may just represent the more
uniform stellar population found in early type galaxies.

\cite{wang97} has also called noted that no SN~Ia has ever been found in
the inner 1~kpc of the center of any spiral galaxy, and suggested again
that this may be a consequence of the strong dust extinction. In
our model the lack of objects in the inner 1 kpc is not unexpected for SNe
arising from a disk population, since the probability of detection in this
region is, indeed, very small (see Figure \ref{f_bestfour}). In our
simulations this is a consequence of the decrease of the disk area element
(2$\pi$r${\it d}$r) toward the galactic center as well as the Shaw effect.

Finally, we have computed the integrated detection efficiency of the
simulation in order to infer the rate of SNe~Ia from the 31 events detected
during the course of the CT survey. For a value of H$_{0}$=65 (M$^{\rm
B}_{1.1}$=-19.258, that obtained in Paper VI) we obtain a rate of
0.21$^{\rm +0.30}_{-0.13}$ SNu in good agreement with the value
0.16$\pm$0.05 SNu recently published by \cite{capellaro97}.  While this is
not a sensitive test of selections effects, it nonetheless lends credence
to our assumptions regarding selection effects in the CT survey.

\acknowledgments

We are grateful to Jill Bechtold, George Jacoby, Bruno Leibundgut, Jim
Liebert, Mark Phillips, Gary Schmidt, Bob Schommer, and Ray White for their
valuable comments.  We thank also an anonymous referee for several useful
suggestions.  MH acknowledges support provided for this work by the
National Science Foundation through grant number GF-1002-97 from the
Association of Universities for Research in Astronomy, Inc., under NSF
Cooperative Agreement No. AST-8947990 and from Fundaci\'{o}n Andes under
project C-12984. Also, MH acknowledges support provided from C\'{a}tedra Presidencial de
Ciencias 1996-1998.
PAP acknowledges support from the National Science
Foundation through CAREER grant AST9501634, from the National Aeronautics
and Space Administration through grant NAG~5-2798, and from the Research
Corporation though a Cottrell Scholarship.

\newpage
\centerline{                             Figure Captions}
\figcaption[Hamuy.fig1.ps]{The complete time sampling for the 60 fields of the
CT survey.  A total of 1,019 photographic plates were obtained. \label{f_obstime}}

\figcaption[Hamuy.fig2.ps]{The distribution of recession velocities, apparent magnitudes,
decline rates, absolute magnitudes, projected angular separations and radial distances
between the 26 ``best-observed'' CT SNe and the host galaxy nuclei. The dashed
histograms show the corresponding distributions for the subsample of
SNe hosted by spiral galaxies. \label{f_ctdata}}

\figcaption[Hamuy.fig3.ps]{The thin solid lines show the contour levels of the KS probability
for the angular separation distribution, for four different values of M$^{\rm B}_{1.1}$.
Overplotted with thick lines are shown the contour levels of the KS probability for
the recession velocity. The contour levels correspond to KS probabilities
of 40\%, 50\%, 60\%, and 70\%. In each panel the $\oplus$ symbol
indicates the values of $\alpha$ and B$_{lim}$ that simultaneously yield the
highest KS probabilities for angular separation and recession velocity. \label{f_simpcont}}

\figcaption[Hamuy.fig4.ps]{(Top panels) The distribution of host galaxy distances and angular
separations between the nuclei of the host galaxies and 3,000 SNe
detected in a Monte Carlo simulation, assuming M$^{\rm B}_{1.1}$=-19.758,
B$_{lim}$=18.3, and $\alpha$=0.15 kpc$^{-1}$. (Bottom panels) Same as
above but for intrinsically less luminous SNe with M$^{\rm B}_{1.1}$=-18.558. \label{f_fourm}}

\figcaption[Hamuy.fig5.ps]{The distribution of recession velocities, apparent magnitudes,
projected angular and radial separations between the nucleus of the host galaxies and
the 26 ``best-observed'' CT SNe.
With solid dots are shown the corresponding distributions (properly scaled)
of 50,000 SNe detected in a simulation performed with
parameters M$^{\rm B}_{1.1}$=-19.258, B$_{lim}$=18.3, and $\alpha$=0.15 kpc$^{-1}$,
a flat luminosity function, and a linear M/$\Delta$m$_{15}$ relation. \label{f_simphist}}

\figcaption[Hamuy.fig6.ps]{(left) The absolute magnitudes of 1,000 SNe
detected in Monte Carlo simulations assuming -19.438$\leq$M${^B}_{MAX}\leq$-18.607
(M$^{\rm B}_{1.1}$=-19.258) and four different selection criteria (see text), plotted
as a function of recession velocity. The dashed diagonal line indicates
the maximum recession velocity up to which each SN type can be found,
the horizontal solid line indicates the average magnitude of the
parent population ($<$M${^B}_{MAX}>$=-19.023), and the horizontal dashed
line corresponds to the average magnitude of the detected SNe. In all
these simulations we assumed B$_{lim}$=18.3, $\alpha$=0.15 kpc$^{-1}$,
$\sigma_{V}$=0, $\sigma_{Blim}$=0, $\sigma_{M}$=0.
(right) The corresponding V/V$_{MAX}$ distributions of the 1,000 detected SNe
in the four cases.  \label{f_vvmax}}

\figcaption[Hamuy.fig7.ps]{(top) The absolute magnitudes of 1,000 SNe detected in a Monte Carlo
simulation assuming M$^{\rm B}_{1.1}$=-19.258 (H$_{0}$=65), B$_{lim}$=18.3,
$\alpha$=0.15 kpc$^{-1}$, $\sigma_{V}$=600 km sec$^{-1}$, $\sigma_{Blim}$=0.3,
and $\sigma_{M}$=0.15, plotted as a function of $\Delta$m$_{15}$(B).
The dashed line corresponds to a linear least-squares fit to
the SNe, and the solid line corresponds to the parent relation.
(second panel from top to bottom) Same as above but for $\sigma_{M}$=0.50.
(third panel from top to bottom) Same as above but for a quadratic parent
M/$\Delta$m$_{15}$ relation.
(bottom panel) Same as above but for a linear M/$\Delta$m$_{15}$ relation
with variable scatter ranging between $\sigma_{M}$=0.15-0.50,
increasing toward larger decline rates. \label{f_bias}}

\figcaption[Hamuy.fig8.ps]{The decline rate and absolute magnitude distributions
of the 26 ``best-observed'' CT SNe.  With solid dots are
shown the corresponding distributions (properly scaled)
of 50,000 SNe detected in a simulation performed with
parameters M$^{\rm B}_{1.1}$=-19.258, B$_{lim}$=18.3, $\alpha$=0.15 kpc$^{-1}$,
a flat luminosity function, and a linear M/$\Delta$m$_{15}$ relation.  \label{f_eff}}

\figcaption[Hamuy.fig9.ps]{The luminosity function of the 26 ``best-observed''
CT SNe (dotted lines) and the corresponding distribution
corrected for selection biases (solid line). The straight line is
an approximate visual fit to the corrected distribution. 
Error bars reflect the uncertainties due to the number of
objects per bin.\label{f_imf}}

\figcaption[Hamuy.fig10.ps]{(top) The solid lines show the fraction 
of detectable events in an elliptical and a spiral galaxy due to the
Shaw effect, as a function of redshift (H$_{0}$=65).
These are analytical calculations where we assume
that the SN radial distribution follows the light distribution of each galaxy.
The dashed line shows how the ratio of elliptical to spiral hosts varies with
redshift.
(bottom) The ratio of elliptical to spiral hosts as
a function of redshift obtained from a Monte Carlo simulation (points)
and the analytical calculation (solid line) for hosts with 
r$_0$=6.7 kpc and r$_{eff}$=6 kpc. The solid line shows the same
calculation, with r$_0$=4 kpc and r$_{eff}$=6 kpc. \label{f_morph}}

\figcaption[Hamuy.fig11.ps]{The apparent magnitudes, absolute magnitudes, and decline rates of
1,000 SNe detected in a simulation performed with parameters
M$^{\rm B}_{1.1}$=-19.258, B$_{lim}$=18.3, $\alpha$=0.15 kpc$^{-1}$, a flat
luminosity function, and a linear M/$\Delta$m$_{15}$ relation, plotted
as a function of recession velocity. The large circles represent the
CT SNe. The vertical line at log(v)=4 (10,000 km sec$^{-1}$)
illustrates the separation of the sample into two bins. For reference, in
the middle panel we have drawn the diagonal dashed line from Figure
\protect\ref{f_vvmax}, that defines the detection limit in an idealized
simulation with $\sigma$$_{M}$=0, $\sigma$$_{V}$=0, and $\sigma$$_{Blim}$=0. \label{f_simpdmdb}}

\figcaption[Hamuy.fig12.ps]{The apparent magnitudes, absolute magnitudes, and
decline rates of 1,000 SNe detected in a simulation performed with
parameters M$^{\rm B}_{1.1}$=-19.258, B$_{lim}$=18.3, $\alpha$=0.15 kpc$^{-1}$,
a flat luminosity function, and a linear M/$\Delta$m$_{15}$ relation,
plotted as a function of projected radial distance between the
SN and the host galaxy nucleus. The large circles represent the
CT SNe. The diagonal line in the top panel
represents the boundary of the central 4 arcsec overexposed
region of the host galaxies where no SNe were discovered. \label{f_simpshaw}}

\figcaption[Hamuy.fig13.ps]{The distribution of recession velocities, apparent magnitudes,
projected angular and radial separations between the nucleus of the host galaxies and
the 26 ``best-observed'' CT SNe.
With solid dots are shown the corresponding distributions (properly scaled)
of 50,000 SNe detected in a simulation performed with
parameters M$^{\rm B}_{1.1}$=-19.258, B$_{lim}$=17.9, $\alpha$=0.175 kpc$^{-1}$,
$\sigma_{Blim}$=0.7, a non-flat luminosity function and
and a non-linear M/$\Delta$m$_{15}$ function. \label{f_bestfour}}

\figcaption[Hamuy.fig14.ps]{Same as for figure \protect\ref{f_bestfour}, but for the decline
rate and absolute magnitude distributions. \label{f_besttwo}}

\figcaption[Hamuy.fig15.ps]{The apparent magnitudes, absolute magnitudes, and
decline rates of 1,000 SNe detected in a simulation performed with
parameters M$^{\rm B}_{1.1}$=-19.258, B$_{lim}$=17.9, $\alpha$=0.175 kpc$^{-1}$,
$\sigma_{Blim}$=0.7, a non-flat luminosity function
and a non-linear M/$\Delta$m$_{15}$ function,
plotted as a function of recession velocity. The large circles
represent the CT SNe. The vertical line at log(v)=4
(10,000 km sec$^{-1}$) illustrates the separation of the sample
into two bins. Typical error bars for the observed quantities
are shown in each panel at log(v)=4.8. \label{f_bestdmdb}}

\figcaption[Hamuy.fig16.ps]{The apparent magnitudes, absolute magnitudes, and
decline rates of 1,000 SNe detected in a simulation performed with
parameters M$^{\rm B}_{1.1}$=-19.258, B$_{lim}$=17.9, $\alpha$=0.175 kpc$^{-1}$,
$\sigma_{Blim}$=0.7, a non-flat luminosity function,
and a non-linear M/$\Delta$m$_{15}$ function,
plotted as a function of projected radial distance between the
SN and the host galaxy nucleus. The large circles represent the
CT SNe. The diagonal line in the top panel
represents the boundary of the central 4 arcsec overexposed
region of the host galaxies where no SNe were discovered.
Typical error bars for the observed quantities
are shown in each panel at a radial distance of 50 kpc. \label{f_bestshaw}}

\figcaption[Hamuy.fig17.ps]{The absolute magnitude distribution
for the synthetic SNe with projected distances from the host galaxy nucleus
larger than 20 kpc (solid lines) and smaller than 10 kpc (dashed line). \label{f_rdist}}

\figcaption[Hamuy.fig18.ps]{(top) The discovery age of 50,000 SNe detected in a simulation performed with
parameters M$^{\rm B}_{1.1}$=-19.258, B$_{lim}$=17.9, $\alpha$=0.175 kpc$^{-1}$,
and $\sigma_{Blim}$=0.7, a non-flat luminosity function,
and a non-linear M/$\Delta$m$_{15}$ function,
plotted as a function of the recession velocity of the host galaxies.
(bottom) Same as above but for the 26 ``best-observed'' CT SNe. \label{f_timedisc}}

\figcaption[Hamuy.fig19.ps]{The color excess of the CT SNe~Ia due to extinction in the
host galaxy as a function of the projected radial distance.
The region bounded by the two horizontal lines corresponds to the location
of the unextinguished SNe. \label{f_dust}}

\clearpage

\begin{deluxetable} {lrrccccrc}
\tablecolumns{9}
\tablenum{1}
\tablewidth{0pt}
\tablecaption{Cal\'{a}n/Tololo Type Ia Supernovae}
\tablehead{
\colhead{SN}  &  
\colhead{ v$_{CMB}$} &
\colhead{ B$_{MAX}$}  & 
\colhead{ $\Delta$m$_{15}$(B)}  & 
\colhead{ M$^{B}$$_{MAX}$}  &
\colhead{ Offset } &
\colhead{ R\tablenotemark{a} $\times$ (H$_{0}$/65)} &
\colhead{ Age } &
\colhead{ Galaxy }\\
\colhead{     } &
\colhead{ [km sec$^{-1}$] } &
\colhead{     } &
\colhead{     } &
\colhead{ -5log(H$_{0}$/65)    } &
\colhead{ [arcsec] } &
\colhead{ [kpc] } &
\colhead{ [days] } &
\colhead{ Type} }
\startdata
 1990T  & 12011 & 17.16 & 1.15  & -19.17 & 24.9  & 22.29   & +10.8 & Sa \nl
 1990Y  & 11633 & 17.70 & 1.13  & -18.56 &  5.1  & 4.43    &  +9.0 & E? \nl
 1990af & 15059 & 17.87 & 1.56  & -18.95 & 10.9  & 12.24   &  -7.3 & SB0 \nl
 1991S  & 16707 & 17.81 & 1.04  & -19.24 & 17.9  & 22.28   &  +8.6 & Sb \nl
 1991U  &  9813 & 16.40 & 1.06  & -19.49 &  6.2  & 4.52    &  +9.1 & Sbc \nl
 1991ag &  4131 & 14.62 & 0.87  & -19.40 & 22.5  & 6.94    &  +3.5 & SBb \nl
 1992J  & 13709 & 17.70 & 1.56  & -18.92 & 16.9  & 17.27   & +12.1 & E/S0 \nl
 1992K  &  3331 & 15.83 & 1.93  & -17.72 & 15.6  & 3.86    &  +9.4 & SBb \nl
 1992O  & 11051 &\nodata&\nodata&\nodata &  7.0  & 5.77    & \nodata & \nodata \nl
 1992P  &  7881 & 16.08 & 0.87  & -19.34 & 10.7  & 6.28    &  -4.0 & SBa \nl
 1992ae & 22448 & 18.62 & 1.28  & -19.07 &  4.5  & 7.55    &  -3.6 & E1? \nl
 1992ag &  7775 & 16.41 & 1.19  & -18.98 &\nodata& \nodata &  -4.5 & S \nl
 1992ai &\nodata&\nodata&\nodata&\nodata &  7.4  & \nodata & \nodata & \nodata \nl
 1992aq & 30287 & 19.45 & 1.46  & -18.89 &  7.5  & 16.87   &  -5.9 & Sa? \nl
 1992bc &  5939 & 15.16 & 0.87  & -19.64 & 16.3  & 7.20    & -12.8 & Sab \nl
 1992bg & 10709 & 16.72 & 1.15  & -19.36 &  6.8  & 5.39    &  +1.8 & Sa \nl
 1992bh & 13517 & 17.70 & 1.05  & -18.89 &  4.1  & 4.10    &  -4.1 & Sbc \nl
 1992bk & 17378 & 18.11 & 1.57  & -19.03 & 24.2  & 31.41   &  +5.9 & E1 \nl
 1992bl & 12889 & 17.36 & 1.51  & -19.13 & 26.7  & 25.62   &  -1.2 & SB0/SBa \nl
 1992bo &  5448 & 15.86 & 1.69  & -18.76 & 72.3  & 29.39   &  -9.9 & E5/S0 \nl
 1992bp & 23673 & 18.41 & 1.32  & -19.40 &  5.6  & 9.800   &  -4.5 & E2/S0 \nl
 1992br & 26324 & 19.38 & 1.69  & -18.66 &  7.2  & 14.22   &  +2.3 & E0 \nl
 1992bs & 18999 & 18.37 & 1.13  & -18.96 &  9.7  & 13.73   &  -1.2 & SBb \nl
 1993B  & 21207 & 18.53 & 1.04  & -19.04 &  5.4  & 8.57    &  -0.1 & SBb \nl
 1993H  &  7441 & 16.84 & 1.69  & -18.45 & 12.4  & 6.86    &  -2.5 & SBb(rs) \nl
 1993M  & 26963 &\nodata&\nodata&\nodata &  3.2  & 6.44    & \nodata & \nodata \nl
 1993O  & 15583 & 17.67 & 1.22  & -19.23 & 16.4  & 19.04   &  -8.3 & E5/S01 \nl
 1993T  & 26175 &\nodata&\nodata&\nodata & 22.9  & 44.71   & \nodata & \nodata \nl
 1993af &  1044 &\nodata&\nodata&\nodata &239.2  & 18.63   & \nodata & \nodata \nl
 1993ag & 15022 & 17.72 & 1.32  & -19.10 &  7.8  & 8.78    & -10.1 & E3/S01 \nl
 1993ah &  8616 & 16.33 & 1.30  & -19.28 &  8.1  & 5.21    &  +3.8 & S02 \nl
\tablenotetext{a}{Projected Galactocentric Distance}
\enddata
\end{deluxetable}

\begin{deluxetable} {rllllll}
\tablecolumns{7}
\tablenum{2}
\tablewidth{0pt}
\tablecaption{B Templates for Type Ia Supernovae}
\tablehead{
\colhead{age}  &  
\colhead{ 1992bc} &
\colhead{ 1991T} &
\colhead{ 1992al} &
\colhead{ 1992A} &
\colhead{ 92bo-93H} &
\colhead{ 1991bg} \\
\colhead{ [days]} &
\colhead{  } &
\colhead{     } &
\colhead{     } &
\colhead{     } &
\colhead{  } &
\colhead{  } }
\startdata
 -14.0 & 2.42:& 1.69:& 3.30:& 4.35:& 5.42:& 7.03:\nl
 -12.0 & 1.61:& 1.02 & 2.10:& 2.77:& 3.18:& 5.00:\nl
 -10.0 & 1.03 & 0.62 & 1.25:& 1.69:& 1.79:& 3.37:\nl
  -8.0 & 0.59 & 0.35 & 0.70:& 0.90:& 0.95 & 2.08:\nl
  -5.0 & 0.25 & 0.14 & 0.20 & 0.30 & 0.34 & 0.79:\nl
  -4.0 & 0.16 & 0.09 & 0.13 & 0.19 & 0.21 & 0.50:\nl
  -3.0 & 0.10 & 0.05 & 0.07 & 0.10 & 0.12 & 0.29:\nl
  -2.0 & 0.06 & 0.02 & 0.03 & 0.04 & 0.06 & 0.14:\nl
  -1.0 & 0.02 & 0.00 & 0.00 & 0.01 & 0.01 & 0.05:\nl
  +0.0 & 0.00 & 0.00 & 0.00 & 0.00 & 0.00 & 0.00 \nl
\nodata&\nodata&\nodata&\nodata&\nodata&\nodata&\nodata \nl
\nodata&\nodata&\nodata&\nodata&\nodata&\nodata&\nodata \nl
\nodata&\nodata&\nodata&\nodata&\nodata&\nodata&\nodata \nl
 +80.0 & 3.62 & 3.46 & 3.73 & 3.80 & 3.96 & 3.58 \nl
 +600.0:& 12.09:& 10.81:& 11.88:& 11.01:& 12.21:& 15.25:\nl
\tablecomments{Values with : denote extrapolations to the data.}
\tablecomments{The template values between days 0 and +80
can be found in Table 1 of Paper VIII.}

\enddata
\end{deluxetable}

\begin{deluxetable} {cccc}
\tablecolumns{4}
\tablenum{3}
\tablewidth{0pt}
\tablecaption{Parameters of the Monte Carlo simulations}
\tablehead{
\colhead{Parameter}  &  
\colhead{Default Value} &
\colhead{Description} &
\colhead{Fixed/Free?} }
\startdata
M$^{B}$$_{1.1}$ & -19.280 & Absolute B magnitude of a SN with $\Delta$m$_{15}$(B)=1.1 & Free  \nl
a$_{0}$ & -19.258 & Zero point of the absolute magnitude-decline rate relation\tablenotemark{a} & Fixed \nl
b$_{0}$ & 0.784 & Slope of the absolute magnitude-decline rate relation\tablenotemark{a} & Fixed \nl
\tablenotetext{a}{M$^{B}$$_{MAX}$ - 5 log(H$_{0}$/65) = a$_{0}$ + b$_{0}$ [$\Delta$m$_{15}$(B) - 1.1]}
$\sigma$$_{M}$ & 0.15 mag & Scatter of the absolute magnitude-decline rate relation\tablenotemark{a} & Fixed \nl
$\sigma$$_{V}$ & 600 km sec$^{-1}$ & Peculiar velocity of host galaxies & Fixed \nl
B$_{lim}$ & 18.30 & B limiting magnitude for SN discoveries & Free \nl
$\sigma$$_{Blim}$ & 0.30 & Scatter in the B limiting magnitude & Fixed \nl
$\alpha$ & 0.15 kpc$^{-1}$ & Inverse of scale length of galaxy disks & Free \nl
separation & 4.0 arcsec & Minimum separation between SN and galaxy nucleus for detection & Fixed \nl
\enddata
\end{deluxetable}

\begin{deluxetable} {ccccccccc}
\tablecolumns{9}
\tablenum{4}
\tablewidth{0pt}
\tablecaption{Rate of Type Ia Supernovae}
\tablehead{
\colhead{N$_{gen}$}  &  
\colhead{N$_{det}$}  &  
\colhead{M${^B}$$_{1.1}$}  & 
\colhead{H$_{0}$}  & 
\colhead{B$_{lim}$}  &
\colhead{$\sigma$$_{Blim}$}  &
\colhead{$\alpha$ } &
\colhead{Radius} &
\colhead{SN Rate} \\
\colhead{      } &
\colhead{      } &
\colhead{      } &
\colhead{[km sec$^{-1}$ Mpc$^{-1}$] } &
\colhead{      } &
\colhead{      } &
\colhead{[kpc$^{-1}$] } &
\colhead{[Mpc]   } &
\colhead{ [SNu]  } } 
\startdata
11538665 &  9000 & -19.258&  65.0&  17.90& 0.70  &  0.175 & 1404.96&  0.21$^{\rm +0.30}_{-0.13}$ \nl
\enddata
\end{deluxetable}

\end{document}